\documentclass[twocolumn,amsmath,amssymb]{revtex4}
\usepackage{amsfonts}
\usepackage[usenames]{color}

\newcommand{\be}{\begin{equation}}
\newcommand{\ee}{\end{equation}}
\newcommand{\bea}{\begin{eqnarray}}
\newcommand{\eea}{\end{eqnarray}}

\begin{document}
\title{On graviton-photon conversions in magnetic environments}
\author{Jai-chan Hwang${}^{1}$, Hyerim Noh${}^{2}$}
\address{${}^{1}$Particle Theory  and Cosmology Group,
         Center for Theoretical Physics of the Universe,
         Institute for Basic Science (IBS), Daejeon, 34126, Republic of Korea
         \\
         ${}^{2}$Theoretical Astrophysics Group, Korea Astronomy and Space Science Institute, Daejeon, Republic of Korea
         }


\begin{abstract}

Graviton-photon conversions in a given external electric or magnetic field, known as the Gertsenshtein mechanism, are usually treated using the four-potential for photons. In terms of the electric and magnetic (EM) fields, however, proper identification of the fields in curved spacetime is important. By misidentifying the fields in Minkowski form, as is often practiced in the literature, we show that the final equation for photon conversion is correct in transverse-tracefree gauge {\it only} for planar gravitational waves in a uniform and constant external field. Even in the former method, to recover the EM fields from the four-potential in curved spacetime, one should properly take into account the metric involved in the relation. By including the metric perturbation in the graviton conversion equation, we show that a magnetic environment can cause tachyonic instability term in gravitational wave equation.

\end{abstract}

\maketitle

%
%
%
\section{Introduction}

The gravitational wave generation mechanism by photons in a uniform and constant external electric or magnetic field is studied by Gertsenshtein \cite{Gertsenshtein-1961}. The opposite process, known as the inverse Gertsenshtein process is also possible and provides potential methods of detecting gravitational waves using electromagnetic means \cite{Lupanov-1967, Boccaletti-1970, Zeldovich-1973, Zeldovich-Novikov-1983}; for recent applications in the high-frequency gravitational wave detection and cosmological constraints, see \cite{Aggarwal-2021, Domcke-2021, Vagnozzi-2022, Ramazanov-2023, Liu-2023}. The original studies used the electric and magnetic (EM) fields for photons (electromagnetic waves), while a more popular formulation uses the four-potential for photons \cite{Raffelt-1988}.

When we use the EM fields for photons, for the correct analysis, it is important to properly identify the EM fields from the field strength tensor based on the covariant decomposition using the observer's four-vector \cite{Moller-1952, Lichnerowicz-1967, Ellis-1973, EM-definition}. This may sound obvious, but in curved spacetime, this fundamental fact is often ignored by a widespread misconception that $F_{ab}$, with the two covariant indices, is independent of the metric. This is not true if $F_{ab}$ is expressed in terms of the EM fields. Missing the metric in $F_{ab}$ directly leads to the homogeneous Maxwell's equations, the same form known in Minkowski space. For a recent occurrence in the context of the Gertsenshtein process, see \cite{Palessandro-Rothman-2023}.

The confusion may arise as $F_{ab}$ expressed in terms of the four-potential is independent of the metric, but {\it not} in terms of the EM fields, see \cite{Berlin-2022, Domcke-2022} for recent cases. Consequently, while the homogeneous Maxwell's equation is identically valid in terms of the potential, it is affected by the metric in terms of the EM fields and becomes nontrivial. If the basic equations are incorrect, the result must be incorrect. Here, we will show that, using a transverse-tracefree gauge for gravitational waves, incorrect analysis happens to give the correct result for a uniform and constant (independent of the space and time coordinate) external magnetic field and a plane gravitational wave which is the case studied in the literature, but not in general, see Sec.\ \ref{sec:wrong}.

The analysis is often made using the four-potential for photons \cite{Raffelt-1988}. The four-potential depends on the electrodynamic gauge choice, whereas the EM fields are gauge-invariant. Besides the gauge dependence, it is not well known that, in curved spacetime, the relation between the potential and EM fields involves the metric perturbation even to the linear order \cite{Hwang-Noh-2023-EM-NL}; in this way, $F_{ab}$ expressed using the EM fields involves metric. To recover the EM fields from the potential, we should take into account the metric (gravitational waves) involved in the relation. Without considering the metric, we recover the wrong result as the one using the EM variables na\"ively defined in special relativistic form, see Sec.\ \ref{sec:potential}. Thus, previous works using the potential are incomplete in this sense.

Our main emphasis in this work is clarifying the above two points (metric dependence of $F_{ab}$ in terms of the EM fields and of the relation between the potential and the EM fields) in the context of graviton-photon conversions. Recently, we have resolved similar issues appearing in the high-frequency gravitational wave detection proposals using electromagnetic means \cite{EM-definition} and in the medium interpretation of gravity in electrodynamics \cite{Hwang-Noh-2023-EM-NL}.

In Secs.\ \ref{sec:graviton} and \ref{sec:photon} we derive equations for the photon-to-graviton conversion and its inverse process, respectively, using both the EM fields and the potential for photons. We consistently consider linear perturbations in both gravitons and photons in a magnetic environment. In Sec.\ \ref{sec:combined} we combine the two processes in two forms. We discuss the result in Sec.\ \ref{sec:discussion}. In Appendix \ref{sec:WG}, we present the photon conversion equation in a general weak gravity situation and compare it with the incorrect one missing the metric.

%
%
%
\section{Graviton conversion}
                                        \label{sec:graviton}

We introduce perturbed metric in Minkowski background
\bea
   g_{ab} = \eta_{ab} + h_{ab}.
\eea
To the linear order, the index of $h_{ab}$ can be raised and lowered using $\eta_{ab}$ and its inverse, and Einstein's equation gives
\bea
   & & h^c_{a,bc} + h^c_{b,ac}
       - h^c_{c,ab} - h^{\;\;\;\;,c}_{ab\;\;\; c}
       - \eta_{ab} ( h^{cd}_{\;\;\;,cd}
       - h^{c,d}_{c\;\;\; d} )
   \nonumber \\
   & & \qquad
       = {16 \pi G \over c^4} \delta T_{ab}.
   \label{GW-eq-ab}
\eea
For derivation, see Eq.\ (6) in \cite{EM-definition}. For gravitational waves, the transverse-tracefree (TT) gauge condition imposes
\bea
   h^j_{i,j} \equiv 0 \equiv h^i_i, \quad
       h_{00} \equiv 0 \equiv h_{0i}.
\eea
The latter two conditions can be regarded as {\it ignoring} non-tensorial perturbations. Under the TT gauge, Eq.\ (\ref{GW-eq-ab}) becomes
\bea
   ( \partial_0^2 - \Delta ) h_{ij}
       = {16 \pi G \over c^4} \delta T_{ij}^{\rm TT},
   \label{GW-eq-ij}
\eea
where
\bea
   T_{ij}^{\rm TT} \equiv P_{ik} T^{k\ell} P_{\ell j}
       - {1 \over 2} P_{ij} P_{k\ell} T^{k\ell}, \quad
       P_{ij} \equiv \delta_{ij} - \hat n_i \hat n_j,
\eea
with $\hat n{}_i$ the propagation direction of a plane gravitational wave with $\hat n{}^i \hat n{}_i \equiv 1$, and $P_{ij}$ is the transverse projection operator \cite{Maggiore-2007}; we have $h_{ij} \hat n^j \equiv 0$.
Considering a plane wave propagating in $z$-direction, thus $\hat {\bf n} = \hat {\bf z} = \hat {\bf x}_3$ and $h_{ij} = h_{ij} (x^0 - z)$, $h_{ij}$ can be expanded using polarization tensors as
\bea
   h_{ij} \equiv h_+ e^+_{ij} + h_\times e^\times_{ij},
\eea
with
\bea
   h_{11} = - h_{22} \equiv h_+, \quad
       h_{12} = h_{21} \equiv h_\times.
\eea

Energy-momentum tensor of the EM fields, in Heaviside unit, is
\bea
   T_{ab} = F_{ac} F_b^{\;\;c}
       - {1 \over 4} g_{ab} F^{cd} F_{cd}.
   \label{Tab-Fab}
\eea
We will consider the coupling between background EM fields and the incoming photons (written as the perturbed EM fields or potential) as the source of gravitational waves. We include the linear order gravitational waves coupled with the background magnetic field. In previous literature, the metric perturbations present in Eq.\ (\ref{Tab-Fab}) are ignored \cite{Raffelt-1988, Ejlli-Thandlam-2019, Fujita-2020, Domcke-2021, Palessandro-Rothman-2023, Ramazanov-2023}. By including the metric contribution it leads to a tachyonic instability term of gravitational waves sourced by the magnetic environment.

The field potential is defined as
\bea
   F_{ab} \equiv A_{b;a} - A_{a;b} = A_{b,a} - A_{a,b},
\eea
and the relation does not involve the metric. In terms of potential the homogeneous Maxwell's equation in Eq.\ (\ref{Maxwell-Fab-2}) is identically satisfied. However, the EM fields depend on the observer's four-vector, $u_a$, and are defined by covariantly decomposing $F_{ab}$ using the four-vector. Using the normal four-vector, $n_a$, corresponding to an Eulerian observer \cite{Smarr-York-1978}, we have \cite{Moller-1952, Lichnerowicz-1967, Ellis-1973}
\bea
   F_{ab} \equiv n_a E_b - n_b E_a - \eta_{abcd} n^c B^d.
\eea
In TT gauge, we have
\bea
   n_a = (-1, 0, 0, 0), \quad n^a = (1, 0, 0, 0),
\eea
and $\eta_{0ijk} = - \sqrt{-g} \eta_{ijk} = - \eta_{ijk}$, and we define $\widetilde A_i \equiv A_i$, $\widetilde B_i \equiv B_i$ and $\widetilde E_i \equiv E_i$ where indices of $\eta_{ijk}$, $A_i$, $B_i$ and $E_i$ are associated with $\delta_{ij}$ and its inverse; if needed, for clarity we use tildes to indicate the covariant quantities.

\subsection{In terms of EM fields}

In terms of the EM fields in the normal frame, in TT gauge, $F_{ab}$ is
\bea
   F_{0i} = - E_i, \quad
       F_{ij} = \eta_{ijk} ( B^k - h^{k\ell} B_\ell ),
   \label{Fab-EB-metric}
\eea
where the gravitational wave appears; to the linear order without TT gauge, see Eq.\ (10) of \cite{EM-definition} and to nonlinear order, see Eq.\ (26) in \cite{Hwang-Noh-2023-EM-NL}.

Now, we decompose the EM fields to the background fields and generated photon as $B_i \rightarrow B_i + b_i$ and $E_i \rightarrow E_i + e_i$. {\it Ignoring} the background $E_i$ compared with $B_i$, we have
\bea
   & & \delta T_{ij} = - B_i b_j - B_j b_i
       + \delta_{ij} B^k b_k
   \nonumber \\
   & & \qquad
       + {1 \over 2} ( h_{ij} B^2
       - \delta_{ij} h^{k\ell} B_k B_\ell ).
\eea
Under the TT operation, we have
\bea
   & &
       \delta T_{ij}^{\rm TT}
       = - B_i b_j - B_j b_i
       + \delta_{ij} ( B^k b_k
       - B^k \hat n_k b^\ell \hat n_\ell )
   \nonumber \\
   & & \qquad
       + 2 \hat n_{(i} ( B_{j)} b^k \hat n_k
       + b_{j)} B^k \hat n_k )
   \nonumber \\
   & & \qquad
       - \hat n_i \hat n_j ( B^k b_k
       + B^k \hat n_k b^\ell \hat n_\ell )
       + {1 \over 2} B^2 h_{ij},
\eea
where $A_{(ij)} \equiv {1 \over 2} (A_{ij} + A_{ji})$. Considering a plane gravitational wave propagating in $z$-direction, we have
\bea
   & & \delta T_{11}^{\rm TT}
       = - \delta T_{22}^{\rm TT}
       = - B_1 b_1 + B_2 b_2
       + {1 \over 2} B^2 h_+,
   \nonumber \\
   & & \delta T_{12}^{\rm TT}
       = - B_1 b_2 - B_2 b_1
       + {1 \over 2} B^2 h_\times.
\eea
Equation (\ref{GW-eq-ij}) gives
\bea
   & & \hskip -.8cm
       ( \partial_0^2 - \partial_z^2
       - {8 \pi G \over c^4} B^2 ) h_+
       = - {16 \pi G \over c^4} (B_1 b_1 - B_2 b_2 ),
   \label{GW-B-1} \\
   & & \hskip -.8cm
       ( \partial_0^2 - \partial_z^2
       - {8 \pi G \over c^4} B^2 ) h_\times
       = - {16 \pi G \over c^4} (B_1 b_2 + B_2 b_1 ).
   \label{GW-B-2}
\eea
Thus, no graviton conversion occurs along parallel (i.e., along $z$-direction) components of the background magnetic field and the impinging photons.

The energy-momentum tensor of electrodynamics contains gravitational waves as well. The metric perturbations coupled with background magnetic field in Eq.\ (\ref{Tab-Fab}) cause what we call the tachyonic instability term in graviton side. In Fourier space with $h_{+,\times} \propto e^{i(\omega t - {\bf k} \cdot {\bf x})}$, 
ignoring the photon source terms, the left-hand sides of Eqs.\ (\ref{GW-B-1}) and (\ref{GW-B-2}) give $\omega^2/c^2 = k^2 - k_B^2$ with $k_B \equiv \sqrt{8 \pi G} B/ c^2$. For a steady $B^2$, we have a solution
\bea
   h_{+,\times} \propto {\rm exp} \Big( \pm
       \sqrt{ {8 \pi G \over c^2} B^2 - c^2 k^2 } t \Big).
\eea
In the large-scale limit with $k \ll k_B$, we have exponential instability with $h_{+,\times} \propto e^{\pm \sqrt{8 \pi G} B t/c}$.

Therefore, in a strong magnetic field sustained for enough scale, gravitational waves go through exponential instability on wavelength scales larger than
\bea
   \lambda_B \equiv {2 \pi \over k_B}
       = 1.4 {1{\rm gauss} \over B} {\rm Mpc}.
   \label{lambda_B}
\eea
For the fast-spinning neutron star with $B \sim 10^{13}$gauss, we have $\lambda_B \sim 4 \times 10^{11}$cm, which is much larger than the radius of a neutron star with $\sim 10^6$cm and the size of the pulsar magnetosphere which may be few orders of magnitude larger than the pulsar radius \cite{Philippov-2022}. Thus, the exponential instability looks not feasible in ordinary astrophysical situations.

In fact, we have a theoretical reason for doubting the realization of such an exponential instability stage. We can estimate
\bea
   & & {L^2 \over \lambda_B^2}
       = {2 \over \pi} {L^2 G B^2 \over c^4}
       \sim {G M \over L c^2} {B^2 \over \varrho c^2}
   \nonumber \\
   & & \qquad
       \sim 5.2 \times 10^{-9}
       \left( {L \over 10^7 {\rm cm}} \right)^2
       \left( {B \over 10^{13} {\rm gauss}} \right)^2
   \nonumber \\
   & & \qquad
       \sim 5.2 \times 10^{-35}
       \left( {L \over 10^2 {\rm cm}} \right)^2
       \left( {B \over 10^5 {\rm gauss}} \right)^2,
   \label{relativistic}
\eea
where we used $M \sim \varrho L^3$ with $M$ and $L$ characteristic mass and length, respectively. Each of the two dimensionless terms in the second evaluation characterizes the relativistic strength of the system. Thus, $L > \lambda_B$ implies the background system is in fully relativistic stage. This means that under this condition we cannot ignore the relativistic gravity of the background magnetic field, thus violating the basic assumption of the graviton-photon conversion analysis which ignore the gravity of the background magnetic field as a hidden assumption. Therefore, the presence of tachyonic instability term is real, but the dominance of this term, so that we have exponential instability stage, is beyond our current treatment. For another theoretical constraint against achieving the exponential instability, see below Eq.\ (\ref{P}).

\subsection{In terms of the potential}

For the background EM fields we have Eq.\ (\ref{Fab-EB-metric}) ignoring the metric. For photons, we have
\bea
   \delta F_{0i} = A_{i,0} - A_{0,i}, \quad
       \delta F_{ij} = A_{j,i} - A_{i,j}.
   \label{Fab-EB-no-metric}
\eea
We use $A_0$ and $A_i$ for photons, i.e., perturbed parts of the potential. Equation (\ref{Tab-Fab}) gives
\bea
   & & \delta T_{ij} = 2 B_{(i} \eta_{j)k\ell} A^{k,\ell}
       - \delta_{ij} B^m \eta_{mk\ell} A^{k,\ell}
   \nonumber \\
   & & \qquad
       + {1 \over 2} ( h_{ij} B^2
       + \delta_{ij} h^{k\ell} B_k B_\ell )
       - 2 h^k_{(i} B_{j)} B_k.
\eea
Under the TT operation, we have
\bea
   & & \delta T_{ij}^{\rm TT}
       = 2 B_{(i} \eta_{j)k\ell} A^{k,\ell}
   \nonumber \\
   & & \qquad
       - 2 \hat n_{(i} \left( B_{j)} \eta_{mk\ell} A^{k,\ell}
       + B_m \eta_{j)k\ell} A^{k,\ell} \right) \hat n^m
   \nonumber \\
   & & \qquad
       + \delta_{ij}
       (B_m \hat n^m \hat n^n \eta_{nk\ell} A^{k,\ell}
       - B^m \eta_{mk\ell} A^{k,\ell} )
   \nonumber \\
   & & \qquad
       + \hat n_i \hat n_j
       (B_m \hat n^m \hat n^n \eta_{nk\ell} A^{k,\ell}
       + B^m \eta_{mk\ell} A^{k,\ell} )
   \nonumber \\
   & & \qquad
       + {1 \over 2} h_{ij} B^2
       + \delta_{ij} h^{k\ell} B_k B_\ell
       - 2 h^k_{(i} B_{j)} B_k
   \nonumber \\
   & & \qquad
       + 2 \hat n_{(i} h^\ell_{j)} B_\ell \hat n_k B^k
       - \hat n_i \hat n_j h^{k\ell} B_k B_\ell.
\eea
Aligning the wave propagation in $z$-direction, we have
\bea
   & & \delta T_{11}^{\rm TT}
       = - \delta T_{22}^{\rm TT}
       = B_1 ( A_{2,3} - A_{3,2} )
   \nonumber \\
   & & \qquad
       - B_2 ( A_{3,1} - A_{1,3} )
       + {1 \over 2} ( B_3^2 - B_1^2 - B_2^2 ) h_+,
   \nonumber \\
   & & \delta T_{12}^{\rm TT}
       = B_1 ( A_{3,1} - A_{1,3} )
       + B_2 ( A_{2,3} - A_{3,2} )
   \nonumber \\
   & & \qquad
       + {1 \over 2} ( B_3^2 - B_1^2 - B_2^2 ) h_\times.
\eea
Equation (\ref{GW-eq-ij}) gives
\bea
   & & \hskip -.8cm
       \left[ \partial_0^2 - \partial_z^2
       - {8 \pi G \over c^4} ( B_3^2 - B_1^2 - B_2^2 )
       \right] h_{+}
   \nonumber \\
   & & \hskip -.8cm
       \qquad
       = {16 \pi G \over c^4} [ B_1 ( A_{2,3} - A_{3,2} )
       - B_2 ( A_{3,1} - A_{1,3} ) ],
   \label{GW-A-1} \\
   & & \hskip -.8cm
       \left[ \partial_0^2 - \partial_z^2
       - {8 \pi G \over c^4} ( B_3^2 - B_1^2 - B_2^2 )
       \right] h_{\times}
   \nonumber \\
   & & \hskip -.8cm
       \qquad
       = {16 \pi G \over c^4} [ B_1 ( A_{3,1} - A_{1,3} )
       + B_2 ( A_{2,3} - A_{3,2} ) ].
   \label{GW-A-2}
\eea
Thus, no graviton conversion occurs along parallel components of the background magnetic field.

The graviton instability term differs from the one in Eqs.\ (\ref{GW-B-1}) and (\ref{GW-B-2}), see below Eq.\ (\ref{h-A-lambda-2}). Relations between the potential and EM fields in the presence of the gravitational waves in TT gauge are \cite{Hwang-Noh-2023-EM-NL}
\bea
   E_i = \partial_i A_0 - \partial_0 A_i, \quad
       B_i = ( \delta_{ij} + h_{ij} ) \eta^{jk\ell}
       \partial_k A_\ell.
   \label{EM-A-TT}
\eea
Notice the gravitational wave involved in the relation. As we consider background EM fields and perturbed photons expressed in terms of either $b_i$-$e_i$ or $A_i$, we have
\bea
   b^i = \eta^{ijk} \partial_j A_k + h^i_j B^j, \quad
       e_i = - A_{i,0},
   \label{b-A}
\eea
where we used $A_0 = 0$, see below Eq.\ (\ref{A_0}). Using this we can show that Eqs.\ (\ref{GW-A-1}) and (\ref{GW-A-2}) satisfies Eqs.\ (\ref{GW-B-1}) and (\ref{GW-B-2}).

The relation between the EM fields and four-potential in general curved spacetime is presented in Eq.\ (89) of \cite{Hwang-Noh-2023-EM-NL}. In the weak gravity limit (i.e., linear perturbations in Minkowski background) without imposing gauge condition, we have
\bea
   & & E_i = ( 1 - {1 \over 2} h^0_0 )
       ( \partial_i A_0 - \partial_0 A_i )
       - h^j_0 ( \partial_i A_j - \partial_j A_i ),
   \nonumber \\
   & & B_i = ( 1 - {1 \over 2} h^\ell_\ell ) \eta_{ijk}
       \partial^j A^k
       + \eta^{jk\ell} h_{ij} \partial_k A_\ell.
   \label{EM-A-linear}
\eea

%
%
%
\section{Photon conversion}
                                        \label{sec:photon}

The inhomogeneous Maxwell's equations give
\bea
   & & \hskip -.9cm
       (\sqrt{-g} F^{ab})_{,b}
       = {1 \over c} \sqrt{-g} J^a
   \nonumber \\
   & & \hskip -.9cm
       \qquad
       = \eta^{ac} \eta^{bd} F_{cd,b}
       + \left( {1 \over 2} h^c_c F^{ab}
       - F^{ac} h^b_c + F^{bc} h^a_c \right)_{,b}.
   \label{Maxwell-Fab-1}
\eea
Notice that to the linear order in metric perturbations, indices of $F_{ab}$ of the last three terms in the right-hand side can be raised and lowered using $\eta_{ab}$ and its inverse, but {\it not} for the first term because $F_{ab}$ with two covariant indices still includes the metric perturbations in terms of EM fields, see Eq.\ (\ref{Fab-EB-metric}). In the following, we {\it ignore} the external source $J_a$. The homogeneous equation is
\bea
   \eta^{abcd} F_{bc,d} = 0,
   \label{Maxwell-Fab-2}
\eea
which also depends on the metric as $F_{ab}$ involves metric perturbation in terms of the EM fields \cite{EM-definition, Hwang-Noh-2023-EM-NL}.

\subsection{In terms of EM fields}

From Eqs.\ (\ref{Maxwell-Fab-1}) and (\ref{Maxwell-Fab-2}) using Eq.\ (\ref{Fab-EB-metric}), we have
\bea
   & & E^i_{\;\;,i} = ( h^{ij} E_j )_{,i},
   \label{Maxwell-1} \\
   & & E^i_{\;\;,0} - \eta^{ijk} \nabla_j B_k
       = ( h^{ij} E_j )_{,0},
   \label{Maxwell-2} \\
   & & B^i_{\;\;,i} = ( h^{ij} B_j )_{,i},
   \label{Maxwell-3} \\
   & & B^i_{\;\;,0} + \eta^{ijk} \nabla_j E_k
       = ( h^{ij} B_j )_{,0}.
   \label{Maxwell-4}
\eea
For the general form without imposing TT gauge, see Eqs.\ (11)-(15) in \cite{EM-definition}. Notice the presence of metric perturbation in {\it all} equations.

We set $B_i \rightarrow B_i + b_i$ and {\it assume} $E_i \ll B_i$, thus $E_i \rightarrow e_i$. From Eqs.\ (\ref{Maxwell-1})-(\ref{Maxwell-4}), we have
\bea
   & & ( \partial_0^2 - \Delta ) b_i
       = ( h_i^j B_j )_{,00}
       - \nabla_i ( h^{jk} B_{j,k} ),
   \label{b-eq-TT} \\
   & & ( \partial_0^2 - \Delta ) e_i
       = \eta_{ijk} \nabla^j ( h^k_\ell B^\ell )_{,0}.
   \label{e-eq-TT}
\eea
Equations to the background order show that vanishing background electric field {\it implies} a uniform and constant magnetic field. If we {\it assume} the background fields are uniform and constant, we have
\bea
   & & ( \partial_0^2 - \Delta ) b_i
       = h_{i,00}^j B_j,
   \\
   & & ( \partial_0^2 - \Delta ) e_i
       = \eta_{ijk} \nabla^j ( h^k_{\ell,0} ) B_\ell.
\eea
Aligning the wave propagation along $z$-direction, we have
\bea
   & & ( \partial_0^2 - \Delta ) b_1
       = h_{+,00} B_1 + h_{\times,00} B_2,
   \label{EM-B-1} \\
   & & ( \partial_0^2 - \Delta ) b_2
       = h_{\times,00} B_1 - h_{+,00} B_2,
   \label{EM-B-2} \\
   & & ( \partial_0^2 - \Delta ) b_3
       = 0,
   \label{EM-B-3}
\eea
and
\bea
   & & ( \partial_0^2 - \Delta ) e_1
       = - h_{\times,z0} B_1 + h_{+,z0} B_2,
   \label{EM-e-1} \\
   & & ( \partial_0^2 - \Delta ) e_2
       = h_{+,z0} B_1 + h_{\times,z0} B_2,
   \label{EM-e-2} \\
   & & ( \partial_0^2 - \Delta ) e_3
       = 0.
   \label{EM-e-3}
\eea
Thus, no photon conversion occurs in the parallel direction of the gravitational waves.

\subsection{In terms of wrong EM fields}
                                        \label{sec:wrong}

In the literature we notice analysis ignoring the metric contributions in $F_{ab}$ with two covariant indices while using the EM fields, see \cite{Berlin-2022, Domcke-2022, Palessandro-Rothman-2023}. Maxwell's equations are sensitive on this omission. As the EM fields are physically measurable quantities by certain specified observers, we do not have the freedom to define these fields arbitrarily. The arbitrarily defined EM fields are often not physically measurable ones by any observer, and this corresponds to the ones defined assuming $F_{ab}$ free from the metric, see \cite{EM-definition, Hwang-Noh-2023-EM-NL}. In this sense, concerning the EM fields we can distinguish the right or wrong definitions. Here, we compare the wrong analysis by {\it missing} the metric in $F_{ab}$ with the correct one in the previous subsection.

We use $\hat E_i$ and $\hat B_i$ to distinguish from the correct EM variables, as
\bea
   F_{0i} \equiv - \hat E_i, \quad
       F_{ij} \equiv \eta_{ijk} \hat B^k.
   \label{Fab-EB-wrong}
\eea
Compared with Eq.\ (\ref{Fab-EB-metric}) we have $\hat E_i = E_i$ but not for $\hat B_i$. The EM fields in Eq.\ (\ref{Fab-EB-metric}) are defined in a covariant way using the normal four-vector as the observer's four-velocity \cite{EM-definition}. Whereas, the ones in Eq.\ (\ref{Fab-EB-wrong}) are defined by simply identifying the components of the field strength tensor $F_{ab}$ with two covariant indices as the EM variables ignoring potential presence of the metric in the relation. In the curved spacetime we proved that the presence of metric in the relation between $F_{ab}$ and EM fields are inevitable for any observer with a four-velocity \cite{Hwang-Noh-2023-EM-NL}.

In TT gauge, using Eq.\ (\ref{Fab-EB-wrong}), Eqs.\ (\ref{Maxwell-Fab-1}) and (\ref{Maxwell-Fab-2}) give
\bea
   & & \hat E^i_{\;\;,i} = ( h^{ij} E_j )_{,i},
   \label{Maxwell-wrong-1} \\
   & & \hat E^i_{\;\;,0} - \eta^{ijk} \nabla_j \hat B_k
       = ( h^{ij} E_j )_{,0}
       + \eta^{ijk} ( B_\ell h^\ell_k )_{,j},
   \label{Maxwell-wrong-2} \\
   & & \hat B^i_{\;\;,i} = 0,
   \label{Maxwell-wrong-3} \\
   & & \hat B^i_{\;\;,0} + \eta^{ijk} \nabla_j \hat E_k
       = 0,
   \label{Maxwell-wrong-4}
\eea
where we have $\hat B_i = B_i$ and $\hat E_i = E_i$ to the background order which is unaffected by gravity.
Comparing these with the correct ones in Eqs.\ (\ref{Maxwell-1})-(\ref{Maxwell-4}), we notice that only Eq.\ (\ref{Maxwell-wrong-1}) remains the same; Eqs.\ (\ref{Maxwell-wrong-3}) and (\ref{Maxwell-wrong-4}) miss the metric perturbations whereas Eq.\ (\ref{Maxwell-wrong-2}) has an additional contribution.

By setting $\hat B_i \rightarrow B_i + \hat b_i$ and $\hat E_i \rightarrow \hat e_i$, from Eqs.\ (\ref{Maxwell-wrong-1})-(\ref{Maxwell-wrong-4}), we have
\bea
   ( \partial_0^2 - \Delta ) \hat b_i
       = \Delta ( h_i^j B_j )
       - \nabla_i ( h^{jk} B_{j,k} ),
   \label{b-eq-TT-wrong}
\eea
and, as we have $\hat e_i = e_i$, equation for $\hat e_i$ remains the same as the correct one in Eq.\ (\ref{e-eq-TT}). The only difference compared with Eq.\ (\ref{b-eq-TT}) is in the first term in the right-hand sides replacing $\partial_0^2$ by $\Delta$. This can be understood because we have $\hat b_i = b_i - h_i^j B_j$ which follows by comparing Eqs.\ (\ref{Fab-EB-metric}) with (\ref{Fab-EB-wrong}), see Eq.\ (\ref{hab-b}) for a case without imposing TT gauge. Therefore, the wrong equations give correct result {\it only if} $( \partial_0^2 - \Delta ) (h_i^j B_j) = 0$ in TT gauge situation; without imposing TT gauge, see Appendix \ref{sec:WG}. This condition is satisfied {\it only} for a uniform and constant $B_i$ and a planar gravitational wave as we show below.

{\it Assuming} a uniform external fields, we have
\bea
   ( \partial_0^2 - \Delta ) \hat b_i
       = \Delta h_i^j B_j.
\eea
Aligning the wave propagation along $z$-direction, we have
\bea
   & & ( \partial_0^2 - \Delta ) \hat b_1
       = h_{+,zz} B_1 + h_{\times,zz} B_2,
   \label{EM-B-wrong-1} \\
   & & ( \partial_0^2 - \Delta ) \hat b_2
       = h_{\times,zz} B_1 - h_{+,zz} B_2,
   \label{EM-B-wrong-2} \\
   & & ( \partial_0^2 - \Delta ) \hat b_3
       = 0.
   \label{EM-B-wrong-3}
\eea
Thus, for a plane wave with $h_{+,\times} = h_{+,\times} (x^0 - z)$, these {\it happen} to coincide with Eqs.\ (\ref{EM-B-1})-(\ref{EM-B-3}). Still, Maxwell's equations in Eqs.\ (\ref{Maxwell-wrong-1})-(\ref{Maxwell-wrong-4}) are wrong because in the presence of gravity the $\hat E_i$ and $\hat B_i$ are not EM fields by any sense (for the gravitational wave in TT gauge it happens that $\hat E_i = E_i$ but still not for $\hat B_i$); these do not have corresponding observer's four-vector, and consequently, no observer can measure such variables as the EM fields \cite{EM-definition, Hwang-Noh-2023-EM-NL}.

In Appendix \ref{sec:WG} we compare the equations in the general weak gravity situation without imposing gauge.

\subsection{In terms of the potential}
                                        \label{sec:potential}

Using the potential, defined as $F_{ab} \equiv \partial_a A_b - \partial_b A_a$, the homogenous Maxwell's equation in (\ref{Maxwell-Fab-2}) is identically valid. Imposing Coulomb gauge, $\nabla \cdot {\bf A} \equiv 0$, Eq.\ (\ref{Maxwell-Fab-1}) for $a = 0$, gives
\bea
   \Delta A_0 = h^{ij} E_{i,j}.
   \label{A_0}
\eea
For uniform background fields or for negligible electric field, the right-hand side vanishes, and we may further {\it set} $A_0 = 0$. For $a = i$, we have
\bea
   ( \partial_0^2 - \Delta ) A_i
       = \eta_{ijk} ( h^j_\ell B_\ell )^{,k},
   \label{EM-A-i}
\eea
where we {\it assumed} $E_i \ll B_i$. Aligning $\hat n_i$ in $z$-direction, we have
\bea
   & & ( \partial_0^2 - \Delta ) A_1
       = h_{\times,z} B_1 - h_{+,z} B_2,
   \label{EM-A-1} \\
   & & ( \partial_0^2 - \Delta ) A_2
       = - h_{+,z} B_1 - h_{\times,z} B_2,
   \label{EM-A-2} \\
   & & ( \partial_0^2 - \Delta ) A_3
       = 0.
   \label{EM-A-3}
\eea
Thus, no photon is generated along the direction of gravitational wave propagation.

Using relations between the potential and EM fields in Eq.\ (\ref{b-A}) we can show that Eqs.\ (\ref{EM-B-1})-(\ref{EM-e-3}) are satisfied by Eqs.\ (\ref{EM-A-1})-(\ref{EM-A-3}). If we ignore the metric in Eq.\ (\ref{b-A}), however, the incorrect equations in (\ref{EM-B-wrong-1})-(\ref{EM-B-wrong-3}) are satisfied by Eqs.\ (\ref{EM-A-1})-(\ref{EM-A-3}) instead. Therefore, in order to recover the physical photons from the potential, we should use Eq.\ (\ref{b-A}) with the metric contribution properly taken into account. However, in a more general context, we cannot recover Eqs.\ (\ref{b-eq-TT}) and (\ref{e-eq-TT}) or Eq.\ (\ref{b-eq-TT-wrong}) from Eq.\ (\ref{EM-A-i}) using Eq.\ (\ref{b-A}) with or without metric perturbations. This implies that these photon conversion equations using the EM fields and the potential are valid {\it only} for a uniform and constant external field which is demanded by ignoring the external electric field, see below Eq.\ (\ref{e-eq-TT}).

%
%
%
\section{Combined equations}
                                        \label{sec:combined}

Now, without losing generality, we align ${\bf B}$ in $y$-$z$ plane, so that $B_x = 0$, $B_y = B \sin{\theta}$, and $B_z = B \cos{\theta}$. Using EM fields for photons, Eqs.\ (\ref{GW-B-1}), (\ref{GW-B-2}), (\ref{EM-B-1})-(\ref{EM-e-3}) give
\bea
   & & ( \partial_0^2 - \partial_z^2
       - {8 \pi G \over c^4} B^2
       ) h_+
       = {16 \pi G \over c^4} B_2 b_2,
   \label{h-b-1} \\
   & & ( \partial_0^2 - \partial_z^2
       - {8 \pi G \over c^4} B^2
       ) h_\times
       = - {16 \pi G \over c^4} B_2 b_1,
   \label{h-b-2} \\
   & & ( \partial_0^2 - \Delta ) b_1
       = B_2 h_{\times,00},
   \label{h-b-3} \\
   & & ( \partial_0^2 - \Delta ) b_2
       = - B_2 h_{+,00},
   \label{h-b-4} \\
   & & ( \partial_0^2 - \Delta ) e_1
       = B_2 h_{+,z0},
   \label{h-b-5} \\
   & & ( \partial_0^2 - \Delta ) e_2
       = B_2 h_{\times,z0}.
   \label{h-b-6}
\eea
Thus, $h_+$ is coupled with $b_2$, and $h_\times$ is coupled with $b_1$, whereas $e_1$ and $e_2$ are generated by $h_+$ and $h_\times$, respectively.

Using the potential for photons, Eqs.\ (\ref{GW-A-1}), (\ref{GW-A-2}), (\ref{EM-A-1}), and (\ref{EM-A-2}) give
\bea
   & & \hskip -.8cm
       \left[ \partial_0^2 - \partial_z^2
       - {8 \pi G \over c^4} ( B_3^2 - B_2^2 )
       \right] h_+
       = {16 \pi G \over c^4} B_2 A_{1,z},
   \label{GW-eq-A-combined-1} \\
   & & \hskip -.8cm
       \left[ \partial_0^2 - \partial_z^2
       - {8 \pi G \over c^4} ( B_3^2 - B_2^2 )
       \right] h_\times
       = {16 \pi G \over c^4} B_2 A_{2,z},
   \label{GW-eq-A-combined-2} \\
   & & \hskip -.8cm
       ( \partial_0^2 - \Delta ) A_1
       = - B_2 h_{+,z},
   \\
   & & \hskip -.8cm
       ( \partial_0^2 - \Delta ) A_2
       = - B_2 h_{\times,z},
\eea
where in Eqs.\ (\ref{GW-eq-A-combined-1}) and (\ref{GW-eq-A-combined-2}) we ignored $A_3$ which is free from the gravitational waves. Thus, $h_+$ is coupled with $A_1$, and $h_\times$ is coupled with $A_2$. Using $A_+ \equiv A_1$ and $A_\times \equiv A_2$, with $\lambda = +, \times$, we have
\bea
   & & \hskip -.8cm
       \left[ \partial_0^2 - \partial_z^2
       - {8 \pi G \over c^4} ( B_3^2 - B_2^2 )
       \right] h_\lambda
       = {16 \pi G \over c^4} B_2 A_{\lambda,z},
   \label{h-A-lambda-1} \\
   & & \hskip -.8cm
       ( \partial_0^2 - \Delta ) A_\lambda
       = - B_2 h_{\lambda,z}.
   \label{h-A-lambda-2}
\eea

Notice that the graviton instability terms in Eqs.\ (\ref{h-b-1})-(\ref{h-b-2}) and Eqs.\ (\ref{GW-eq-A-combined-1})-(\ref{GW-eq-A-combined-2}) depend on whether we use the potential or EM fields for photons. This dependence is natural as the relation between the potential and EM fields involves the metric. When we ignore the photon source, Eqs.\ (\ref{h-b-1})-(\ref{h-b-2}) are the correct forms for gravitational waves immersed in external magnetic field. This is because when we ignore photons as the electromagnetic waves we should translate the potential to the EM fields.

From dimensional argument, the conversion rate (energy ratio) of the photon to the gravitational wave can be estimated from Eqs.\ (\ref{h-b-1}) and (\ref{h-b-2}) as \cite{Gertsenshtein-1961, Zeldovich-1973, Zeldovich-Novikov-1983}
\bea
   & & P_{\gamma \rightarrow g}
       \equiv {E_{\rm GW} \over E_{\rm EM}}
       = {c^2 \over 8 \pi G} { |\dot h_+|^2 + |\dot h_\times|^2
       \over |e_1|^2 + |e_2|^2 + |b_1|^2 + |b_2|^2}
   \nonumber \\
   & & \qquad
       \sim {c^2 \over 16 \pi G} { |\dot h|^2
       \over |b|^2}
       \sim {16 \pi G \over c^4} B^2 \sin^2{\theta} L^2,
\eea
where $L$ is the conversion length (the length of background magnetic field), and we used \cite{MTW-1973, Landau-Lifshitz-1975}
\bea
    T^{\rm GW}_{00}
        = {c^4 \over 32 \pi G} h^{ij}_{\;\;\;,0} h_{ij,0}
        = {c^2 \over 16 \pi G} ( |\dot h_+|^2
        + |\dot h_\times|^2 ),
    \label{T00}
\eea
and the energy density of the EM fields $\mu_{\rm EM} = {1 \over 2} ( E^2 + B^2 )$. The inverse conversion rate ($P_{g \rightarrow \gamma} \equiv E_{\rm EM} / E_{\rm GW}$) can be estimated similarly from Eqs.\ (\ref{h-b-3})-(\ref{h-b-6}), and the result is the same as above \cite{Zeldovich-Novikov-1983}. Using $\lambda_B$ (the critical scale for graviton instability) in Eq.\ (\ref{lambda_B}), we have
\bea
   P_{\gamma \rightarrow g}
       \sim P_{g \rightarrow \gamma}
       \sim 8 \pi^2 \sin^2{\theta} (L/\lambda_B)^2.
   \label{P}
\eea
Thus, the graviton and photon conversion rates smaller than equality demands violation of the instability condition $L > \lambda_B$. As mentioned below Eq.\ (\ref{relativistic}), this instability condition implies violation of our hidden assumption ignoring the gravity of the background magnetic field system. Solutions of Eqs.\ (\ref{h-A-lambda-1}) and (\ref{h-A-lambda-2}) are studied in \cite{Ejlli-Thandlam-2019, Ejlli-2020, Fujita-2020, Palessandro-Rothman-2023}.

%
%
%
\section{Discussion}
                                        \label{sec:discussion}

We derived graviton-photon conversion equations in a given external magnetic field adopting TT gauge for gravitational waves. We presented two forms of equations for photons, using EM fields and the potential. The photon conversion equation using the EM fields is new in the sense that previous work in \cite{Palessandro-Rothman-2023} used incorrect definitions of the EM fields. The wrong equations happen to coincide with the correct ones only for a plane gravitational wave in a uniform and constant external magnetic field. For equations using the potential, we pointed out that to recover the EM fields, we should consider metric dependence between the potential and EM fields properly. We showed that by missing the metric, one recovers the wrong equations mentioned above.

Here, we considered linear perturbations in both gravitons and photons consistently. All previous works ignored the metric contribution present in the source term (energy-momentum tensor) of the graviton conversion equation. The metric perturbation is important to include, and it leads to a tachyonic instability term of gravitational waves in the magnetic environment. We also showed that the realization of instability is theoretically not feasible in our formulation ignoring the gravity of the background magnetic field. As the relation between the potential and EM fields depends on the metric, the graviton instability term also depends on using either the potential or EM fields; see below Eq.\ (\ref{h-A-lambda-2}).

The gravity coupling with electrodynamics is often expressed in the following Lagrangian density
\bea
   {\cal L}_{\rm EM}
       = - {1 \over 4} \eta^{ac} \eta^{bd} F_{ab} F_{cd}
       + {1 \over 2} h_{ab} T^{ab},
\eea
which follows from ${\cal L}_{\rm EM} = - {1 \over 4} \sqrt{-g} F^{ab} F_{ab}$ expanded to the linear order in metric perturbations; $g = {\rm det}(g_{ab})$ and $T_{ab}$ is the electromagnetic energy-momentum tensor in Eq.\ (\ref{Tab-Fab}). The second term is often mentioned as the interaction Lagrangian \cite{Boccaletti-1967}.

However, it is important to notice the presence of metric coupling even in the first term; $F_{ab}$ with two covariant indices involves metric when expressed in terms of the EM fields \cite{EM-definition, Hwang-Noh-2023-EM-NL}, see Eq.\ (\ref{Fab-EB-metric}) in our case. Metric disappears if we express $F_{ab}$ in terms of the four-potential, but in that case, the relation between the potential and the EM fields involves the metric again \cite{Hwang-Noh-2023-EM-NL}, see Eqs.\ (\ref{EM-A-TT}) and (\ref{EM-A-linear}). The main point of the present work is clarifying these two points, which were often neglected in the literature: the presence of metric in (i) $F_{ab}$ and (ii) the relation between the potential and EM fields.

Although the photon conversion equations happen to coincide for a uniform and constant magnetic field, spatially non-uniform magnetic field configurations are common in astrophysical situations. The examples are pulsar magnetosphere, magnetic fields associated with accretion disks around compact objects, extragalactic jets, galaxies and galaxy clusters, large-scale structure, etc. \cite{Zeldovich-Ruzmaikin-Sokoloff-1983}. It is rather difficult to imagine a uniform and constant magnetic field configuration except as approximations.

Even in experimental design for high-frequency gravitational wave detection, one can easily imagine curved magnetic field configurations \cite{Domcke-2022} where proper analyses are demanded. Although based on incorrect Maxwell's equations, gravitational wave detection mechanisms using axion search experiments were proposed in \cite{Berlin-2022, Domcke-2022}. In axion search experiments using the magnetic field, a spatial or temporal variation in the magnetic field is known to enhance the conversion probability \cite{Seong-2023}.

The correct Maxwell's equations in the normal frame are no more complicated than the incorrect ones to handle. Although the results happen to coincide in a simple external field configuration, correct analysis is essential for proper understanding. For a non-uniform or time-varying background field, as the wrong equations differ from the correct ones in a non-trivial way, we have no reason to expect the accidental coincidence in the constant and uniform field case maintained. The different terms between the two cases are in the same order as the correct ones.

Here, we derived the graviton-photon conversion equations adopting the TT gauge for the gravitational waves. In Laboratory experiments located on the Earth or its surroundings, the Fermi normal coordinate (FNC) may suit than the TT gauge \cite{Maggiore-2007}. Such attempts were previously made in \cite{Berlin-2022, Domcke-2022}, but based on the wrong Maxwell's equations by using Eq.\ (\ref{Fab-EB-wrong}), see \cite{EM-definition}. It is far from clear that the coincidence in the uniform and constant field found in the TT gauge can be maintained in the FNC.

The Euler-Heisenberg (EH) QED correction in Maxwell's equations is often included in the graviton-photon conversion analysis \cite{Raffelt-1988, Ramazanov-2023}. The EH correction also has its own metric perturbations, which are always ignored in the literature. Neglecting such a contribution due to its expected small nature (small EH correction times small metric perturbation is negligible in ordinary situations) is different from neglecting the metric in the field identification, which is of the same order as the metric in the interaction term in the graviton-photon conversions.

%
%
%
\section*{Acknowledgments}

We thank Drs.\ Sang Hui Im, Chan Park and Seokhoon Yun and Professor Hyun Kyu Lee for useful discussion. We wish to thank two referees for their critical questions and constructive suggestions. H.N.\ was supported by the National Research Foundation (NRF) of Korea funded by the Korean Government (No.\ 2018R1A2B6002466 and No.\ 2021R1F1A1045515). J.H.\ was supported by IBS under the project code, IBS-R018-D1, and by the NRF of Korea funded by the Korean Government (No.\ NRF-2019R1A2C1003031).

\appendix
\section{Photon conversion by weak gravity}
                                        \label{sec:WG}

Here, we present the photon conversion equation in a {\it general} weak gravity environment (including gravitational waves without fixing TT gauge), and compare with the incorrect one ignoring the effect of gravity in $F_{ab}$.

Without imposing the gauge condition, thus using general $h_{ab}$, Maxwell's equations in the normal frame valid to linear order metric perturbation are derived in Eqs.\ (11)-(15) of \cite{EM-definition}. We {\it ignore} the four-current. From the Maxwell's equations, setting $B_i \rightarrow B_i + b_i$ and $E_i \rightarrow e_i$, thus ignoring the background $E_i$, we can derive
\bea
   & & \hskip -.9cm
       ( \partial_0^2 - \Delta ) b^i
       = ( h^{ij} B_j - {1 \over 2} h^j_j B^i )_{,00}
       + {1 \over 2} \Delta ( h_0^0 B^i )
   \nonumber \\
   & & \hskip -.9cm
       \qquad
       - [ h^{jk} B_k - {1 \over 2} ( h^k_k - h^0_0 ) B^j
       ]^{,i}_{\;\;j}
       + ( h^j_0 B^i - h^i_0 B^j )_{,0j},
   \label{b-eq-general} \\
   & & \hskip -.9cm
       ( \partial_0^2 - \Delta ) e^i
       = \eta^{ijk} \nabla_j \big\{
       ( h_0^\ell B_k - h_{0k} B^\ell )_{,\ell}
   \nonumber \\
   & & \hskip -.9cm
       \qquad
       + [ h_k^\ell B_\ell
       - {1 \over 2} ( h^\ell_\ell - h^0_0 ) B_k ]_{,0}
       \big\}.
   \label{e-eq-general}
\eea
In TT gauge, we recover Eqs.\ (\ref{b-eq-TT}) and (\ref{e-eq-TT}). To the background order, ignoring the background electric field demands a uniform and constant magnetic field.

Maxwell's equations using the wrong EM fields defined using Eq.\ (\ref{Fab-EB-wrong}) are derived in Eqs.\ (19)-(23) of \cite{EM-definition}. Similarly, we can derive
\bea
   & & \hskip -.4cm
       ( \partial_0^2 - \Delta ) \hat b^i
       = \Delta [ h^{ij} B_j
       - {1 \over 2} ( h^j_j - h^0_0 ) B^i ]
   \nonumber \\
   & & \hskip -.4cm
       \qquad
       - [ h^{jk} B_k - {1 \over 2}
       ( h^k_k - h^0_0 ) B^j ]^{,i}_{\;\;j}
       + ( h^j_0 B^i - h^i_0 B^j )_{,0j},
   \label{b-eq-general-wrong} \\
   & & \hskip -.4cm
       ( \partial_0^2 - \Delta ) \hat e^i
       = \eta^{ijk} \big\{
       ( \partial_0^2 - \Delta ) ( h_{0j} B_k )
       + ( h_0^\ell B_k - h_{0k} B^\ell )_{,\ell j}
   \nonumber \\
   & & \hskip -.4cm
       \qquad
       + [ h_k^\ell B_\ell
       - {1 \over 2} ( h^\ell_\ell - h^0_0 ) B_k ]_{,0j}
       \big\}.
   \label{e-eq-general-wrong}
\eea
In TT gauge, we recover Eqs.\ (\ref{b-eq-TT-wrong}) and (\ref{e-eq-TT}). Compared with Eqs.\ (\ref{b-eq-general}) and (\ref{e-eq-general}), the only difference appears in the first two terms in the right-hand sides, with $\partial_0^2$ replaced by $\Delta$. This can be understood because, from Eq. (10) or (18) of \cite{EM-definition}, we have
\bea
   b_i = \hat b_i + h^j_i B_j - {1 \over 2} h^j_j B_i, \quad
   \hat e_i = e_i + \eta_{ijk} h^j_0 B^k.
   \label{hab-b}
\eea
In \cite{EM-definition} this relation between variables was suggested to translate the wrong analysis to the correct one.

%
%


\end{document}